\documentclass[twocolumn,10pt]{article}
\pdfoutput=1

\usepackage[letterpaper,margin=2.25cm]{geometry}
\usepackage[T1]{fontenc}
\usepackage{newtxtext,newtxmath}
\usepackage{amsmath}
\usepackage{bm}
\usepackage{graphicx}
\usepackage{booktabs}
\usepackage{tabularx}
\newcolumntype{L}{>{\raggedright\arraybackslash}X}
\usepackage{float}
\usepackage{enumitem}
\usepackage{cite}
\usepackage{caption}
\usepackage{url}
\usepackage{xurl}
\usepackage[colorlinks=false,pdfborder={0 0 0}]{hyperref}
\usepackage{balance}

\newcounter{lemmacount}

\begin{document}

\twocolumn[{%
\begin{center}
{\large\bfseries Space Postman and ATLAS: Enabling Startup Participation in the Orbital Logistics Economy\par}
\vskip 1.5em
{\normalsize
\textbf{Omer Burak Iskender}\textsuperscript{1,*},\
\textbf{Kaan G\"{o}kt\"{u}rk}\textsuperscript{2}\par}
\vskip 1.5em
{\normalsize\textsuperscript{1}\,\textit{Nanyang Technological University, Singapore.} E-mail: \href{mailto:iske0001@e.ntu.edu.sg}{iske0001@e.ntu.edu.sg}\par}
\vskip 0.5em
{\normalsize\textsuperscript{2}\,\textit{Istanbul Technical University, T\"{u}rkiye.} E-mail: \href{mailto:gokturkk19@itu.edu.tr}{gokturkk19@itu.edu.tr}\par}
\vskip 0.5em
{\normalsize\textsuperscript{*}\,Corresponding author\par}
\vskip 1em
{\small\itshape Preprint of a paper accepted for presentation at the 77th International Astronautical Congress (IAC 2026), Antalya, T\"{u}rkiye, 5--9 October 2026.\par}
\end{center}
\vskip 1em
\begin{center}
\textbf{Abstract}
\end{center}
\noindent
Debris removal, in-orbit servicing and refuelling, last-mile delivery, in-space recycling and constellation deployment are separate businesses built on one planning problem: a single spacecraft must visit many targets whose orbital planes drift apart under Earth's oblateness, within a propellant tank and a deadline. Solving it well takes orbital dynamics and optimisation expertise that is rarely the core business of payload, tug or constellation companies; this keeps young operators out of orbital logistics or leaves them with costly hand-made schedules. This paper presents Space Postman, a planning service that turns a target list, a spacecraft and a time budget into a priced and, for rendezvous tours, verified mission plan, and ATLAS, a tug concept whose payload bay takes debris kits, client spacecraft or propellant. As most start-ups fail when the money runs out, the offer starts with software and adds the vehicle later. Two case studies show what it changes. For a notional follow-on to the DEXTER project, which develops ways to turn scrap aluminium into propellant, it plans a tour of ten high-risk upper stages that share one orbital shell: waiting alone cannot close the tour, an even split of the time costs a third more propellant, and time is the currency of the mission. For a tug that releases eight clients from one rideshare, the delivery order decides whether the tank suffices, and the plan prices each delivery by the plane it must reach. The planning engine reproduces the published winning solution of a public debris-removal benchmark at that solution's fidelity and, given the same launches, plans cheaper trajectories for most of them: an independent check for operators and investors. A calibrated planner paired with a payload-agnostic tug offers new operators a practical way into orbital logistics, provided its numbers are read at the fidelity at which they were obtained.
\vskip 2.5ex
}]

\section*{Abbreviations}\nopagebreak
\noindent\small
\begin{tabbing}
\textbf{DEXTER}\hspace{0.5em}\=\kill
\textbf{ATLAS}  \> Autonomous Tug for Logistics and Servicing \\
\textbf{DEXTER} \> Debris Extraction Tools for Extra-terrestrial \\
                \> Recycling (Horizon Europe project) \\
\textbf{ESA}    \> European Space Agency \\
\textbf{LEO}    \> Low Earth Orbit \\
\textbf{LTAN}   \> Local Time of the Ascending Node \\
\textbf{MEUR}   \> Million Euro \\
\textbf{SSO}    \> Sun-Synchronous Orbit \\
\textbf{TRL}    \> Technology Readiness Level \\
\textbf{VC}     \> Venture Capital
\end{tabbing}
\normalsize

\section{Introduction}\label{sec:intro}

Several businesses in low Earth orbit (LEO), many of them young, depend
on the same kind of mission. A tug that releases rideshare clients into their target orbits, a
servicer or tanker that calls on several client satellites, a remover
that collects derelict objects, a recycler that harvests them for raw
material, and a constellation operator that spreads one launch over
several orbital planes all fly one spacecraft to many destinations in
turn, within a propellant tank and a
deadline~\cite{g_barato2024,g_luu2022,g_zhang2014,g_liou2011,g_palmer2025,g_mcgrath2020}.
What these businesses share, and what this paper is about, is the
planning problem underneath them.

\begin{figure}[!htb]
\centering
\includegraphics[width=\columnwidth]{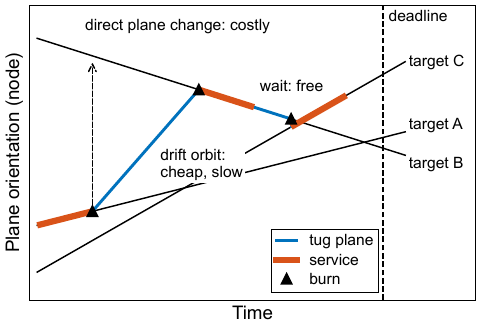}
\caption{The planning problem behind many orbital-logistics businesses
(schematic). The planes of targets A, B and C turn at different rates.
Changing plane at once is costly; a drift orbit turns the tug's plane
slowly for little propellant; waiting is free when a target's plane comes
to the tug. Every choice costs time, and the tour must end before a
deadline with the propellant in one tank.}
\label{fig:problem}
\end{figure}

Figure~\ref{fig:problem} draws that problem. In LEO the dominant cost of
a multi-target mission is the change of orbital plane between targets:
at 800~km a plane change of ten degrees costs about
1.3~km/s~\cite{iskender_iac_c1}. Earth's oblateness turns every plane at
a different rate, so the angle between two planes changes with time. Waiting,
or parking for a while on a \emph{drift orbit} that turns at another
rate, can close the gap for little propellant, but it costs time. The
price of every leg therefore depends on when it is flown, the time spent
on one leg changes the geometry of every later leg, the visiting order
matters, and the tank and the deadline bound the whole tour. In the
language of operations research this is a moving-target, time-dependent
routing problem~\cite{g_bourjolly2006,b_izzo2015}.
Exploiting the drift of the planes is established practice, from
multi-debris tour design and constellation deployment to recent
servicing studies~\cite{g_cerf2016,g_mcdowell2020,g_apa2026,g_ciccotti2026,g_leonardi2026}.

Solving it well is a barrier to entry. It takes orbital dynamics,
combinatorial search and trajectory optimisation, skills that are rarely
the core business of a payload, tug or constellation company. The methods
exist, but in the research
literature~\cite{b_simoes2017,g_zhang2024,g_bannach2026,g_ciccotti2026},
and turning them into a plan that an operator can quote, fly and defend
still takes specialist effort. A newcomer can build that capability, hire
it as consultancy, or schedule by hand. Our case studies show what the
last option costs: an even split of the time budget costs a third more
propellant on a debris tour, a hand-sorted delivery order almost a
quarter more on a tug, and the order in which a manifest happens to be
written does not fit in the tank at all
(Secs.~\ref{sec:dexter}--\ref{sec:value}).

This paper presents two products aimed at that barrier.
\textbf{Space Postman} is a planning service in development: an operator uploads its
targets, its spacecraft and its time budget, and receives a mission plan
with the order, the dates, the velocity change ($\Delta V$) and
propellant of every leg, and, for rendezvous tours, a check at higher
fidelity. \textbf{ATLAS} is a tug concept whose payload bay takes debris
kits, client spacecraft or propellant, so that one vehicle and one
planner serve several of the businesses above. The planning engine and its
evidence are the subject of a companion technical paper at this
congress~\cite{iskender_iac_c1} and of a journal
manuscript~\cite{iskender_journal2026}. The companion paper tests the
planner on three cases: the debris-removal benchmark of ESA's ninth
Global Trajectory Optimisation Competition (GTOC9), a notional tour of
derelict rocket bodies that would put the harvesting developed by the
DEXTER recycling project to work in orbit, and a multi-client tug. This paper takes the business view of the last two and
uses the first only as a credibility check; it repeats none of the
companion's tables or figures.

Section~\ref{sec:market} maps the market and Section~\ref{sec:startups}
the start-up's constraints, Sections~\ref{sec:postman}
to~\ref{sec:trust} present the offer and its credibility,
Sections~\ref{sec:dexter} to~\ref{sec:value} the case studies and
their value, and Section~\ref{sec:roadmap} the roadmap.

\section{One Problem, Many Businesses}\label{sec:market}

The potential customer base is large: ESA counts about 15\,900
functioning satellites in orbit and about 47\,000 regularly tracked
objects (July 2026)~\cite{b_esa_discosweb}.
Table~\ref{tab:market} maps the operators whose vehicles or missions fit
the problem of Fig.~\ref{fig:problem}, by segment, as their announcements stood in September 2026. Three observations shape the
business case.

\begin{table*}[t]
\caption{Operators and projects in the segments that share the
multi-target planning problem, as announced up to 29~September~2026.
Several?: whether one mission visits, serves or deploys to several
targets or orbits.}
\label{tab:market}
\centering
\small
\begin{tabularx}{\linewidth}{@{}>{\hsize=0.95\hsize}L>{\hsize=0.95\hsize}L>{\hsize=1.45\hsize}L>{\hsize=0.65\hsize}L@{}}
\toprule
\textbf{Operator (country)} & \textbf{Vehicle or mission} & \textbf{Status, September 2026} & \textbf{Several?} \\
\midrule
\multicolumn{4}{@{}l}{\emph{Last-mile delivery and orbital transfer}} \\
D-Orbit (IT)~\cite{m_dorbit_origin,m_dorbit_summer,m_dorbit_ion} & ION Satellite Carrier & flying since 2020; 23rd mission July 2026 & yes, a separate orbit per client \\
Exotrail (FR)~\cite{m_exotrail_spacevan} & spacevan & flying since November 2023 & not stated \\
Impulse Space (US)~\cite{m_impulse_mira,m_impulse_helios} & Mira; Helios & Mira: three missions since 2023; Helios: first launch 2027 & not stated \\
Momentus (US)~\cite{m_momentus_missions,m_momentus_vr7} & Vigoride & first flight 2022; Vigoride-7 in orbit since March 2026 & not stated \\
\midrule
\multicolumn{4}{@{}l}{\emph{Servicing, life extension and refuelling}} \\
Northrop Grumman (US)~\cite{m_ng_spacelogistics,m_ng_mrv} & MEV-1, MEV-2; robotic vehicle and pods & MEVs docked 2020--21, MEV-1 to a second client 2025; robotic vehicle launched July 2026 (all GEO) & yes, in sequence \\
Starfish Space (US)~\cite{m_nasa_sspicy,m_viasat_otter} & Otter & first Otter, to inspect several derelicts, launch from October 2026 & planned \\
Orbit Fab (US)~\cite{m_orbitfab_rafti,m_orbitfab_raven} & Tenzing; RAVEN, NEST & depot flown 2021; RAVEN shuttle and NEST depot planned for 2030 & planned \\
\midrule
\multicolumn{4}{@{}l}{\emph{Debris inspection and removal}} \\
Astroscale (JP, UK)~\cite{m_astroscale_adrasj,m_astroscale_adrasj_end,m_astroscale_adrasj2} & ADRAS-J; ADRAS-J2 & 15-m approach to a derelict stage 2024, de-orbiting since March 2026; capture mission ADRAS-J2 in development & no, one each \\
ClearSpace (CH, UK)~\cite{m_esa_clearspace1,m_clearspace_clear} & ClearSpace-1; CLEAR & ClearSpace-1 (ESA): launch planned 2029; CLEAR (UK): design phase & CLEAR: two \\
\midrule
\multicolumn{4}{@{}l}{\emph{In-space recycling and propulsion}} \\
DEXTER (EU)~\cite{g_dexter2025} & EIC Pathfinder project & laboratory and ground work, 2025--2029 & case study: Sec.~\ref{sec:dexter} \\
Magdrive (UK)$^{*}$~\cite{m_magdrive_rogue_iod} & metal-propellant thrusters & first in-orbit demonstration 2025--26; DEXTER partner & -- \\
\midrule
\multicolumn{4}{@{}l}{\emph{Constellation deployment}} \\
SpaceX (US)~\cite{g_mcdowell2020} & Starlink & batches held low for 35 and 70~days to reach planes $20^\circ$ and $40^\circ$ apart & yes, by precession \\
\midrule
\multicolumn{4}{@{}l}{\emph{Planning software and tug concept}} \\
Final Proximity (TR, SG)$^{*}$~\cite{m_fprox} & guidance software; ATLAS & concept & -- \\
\bottomrule
\end{tabularx}
\par\vspace{3pt}
\begin{minipage}{\linewidth}\footnotesize
$^{*}$Relationship with the authors disclosed in the text. Status as
stated on the cited pages, most of them the operators' pages; trade press
only where no primary source was found. GEO: geostationary orbit. The
table is a map of the segments, not a complete census.
\end{minipage}
\end{table*}

First, flight heritage sits at the two ends of the problem. Carriers
that can release each rideshare client into its target orbit fly
routinely~\cite{m_dorbit_ion}, and in geostationary orbit one servicer
has already docked with two clients in turn~\cite{m_ng_spacelogistics}.
In LEO, multi-target servicing, removal and recycling tours are still
demonstrations or plans, and the removal missions in the table have one
target each, apart from the two-object CLEAR design. Second, every segment meets the problem of
Fig.~\ref{fig:problem} as soon as one mission must reach more than one
orbital plane in LEO: a tug whose clients want different local times, a
servicer or remover with several clients or derelicts, a recycler
collecting several stages, and a constellation that fills several planes
from one launch. Third, the segments differ in what happens at each stop
and in whether the target must be met or only its orbit reached; this
changes the mass accounting and the required fidelity, not the core of
the plan, so one planning engine can address them. The companion paper shows this for removal,
recycling and delivery~\cite{iskender_iac_c1}; servicing and refuelling,
with propellant transfers and return legs, are not yet demonstrated.

General-purpose mission-analysis tools, including the free General
Mission Analysis Tool (GMAT)~\cite{b_gmat}, model and target the transfers
that an analyst specifies; to our knowledge, choosing the order and the
timing of a many-target tour is left to the analyst. That choice is the
gap Space Postman is aimed at.

\emph{Disclosure.} The first author is with Final Proximity Space
Systems, which develops guidance and multi-rendezvous optimisation
software and the ATLAS concept and supports this work~\cite{m_fprox}.
Magdrive is a DEXTER
partner~\cite{g_dexter2025} and employs two co-authors of the companion
technical paper~\cite{iskender_iac_c1}. DEXTER is treated as a case study,
not as a customer, and neither it nor Magdrive is presented as a customer;
no DEXTER or Magdrive performance figure is used.

\section{Why a Start-Up Needs a Lean Entry}\label{sec:startups}

\begin{figure*}[t]
\centering
\includegraphics[width=\textwidth]{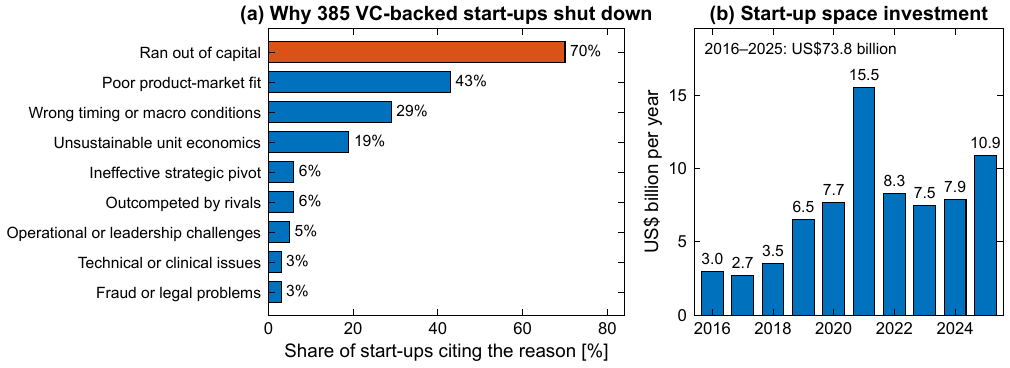}
\caption{Why a start-up needs a lean entry. (a)~Reasons given for the
shutdown of 385 VC-backed start-ups from 2023 to early 2026, several per
company~\cite{m_cbinsights2026}. (b)~Investment in space start-ups
(hardware, launch and satellite operations) per
year~\cite{m_brycetech2026}.}
\label{fig:startups}
\end{figure*}

Most young companies do not fail for want of a product but for want of
money. Of 385 venture-backed start-ups that shut down from 2023 to early
2026 and whose reasons could be identified, 70\% ran out of capital
(Fig.~\ref{fig:startups}a), a median of 22~months after their last
raise~\cite{m_cbinsights2026}; running out of capital is usually the
final cause rather than the root one, but it is the one that ends the
company. Across all sectors, about half of the US private-sector
establishments opened in March~2015 were still operating five years
later, and about a third after ten~\cite{m_bls_bed_survival}.

Space adds capital intensity and a cycle. Investment in space start-ups
totalled US\$73.8~billion from 2016 to 2025; it peaked at 15.5~billion
in 2021 and then nearly halved (Fig.~\ref{fig:startups}b)~\cite{m_brycetech2026}.
When the money stopped, hardware companies were caught: Vector Launch
filed for bankruptcy in 2019 after an investor withdrew its
support~\cite{m_vector2019}, Virgin Orbit in 2023 after failing to secure
new funding~\cite{m_virginorbit2023}, the tug operator Momentus ended 2024
with US\$1.6~million in cash and a going-concern
doubt~\cite{m_momentus10k2024}, and Orbex entered administration in 2026
after its fundraising and merger efforts had failed~\cite{m_orbex2026}.
Investors had long favoured data and imagery start-ups over rocket makers
and hardware suppliers for their quicker path to revenue, although by
2024 more money was going to capital-intensive
ventures~\cite{m_businessreporter2024}. For a newcomer in orbital
logistics this argues for an entry that earns revenue before it needs a
vehicle; Sec.~\ref{sec:value} returns to it.

\section{Space Postman: Planning as a Service}\label{sec:postman}

\subsection{What the customer supplies and receives}

\begin{figure*}[t]
\centering
\includegraphics[width=\textwidth]{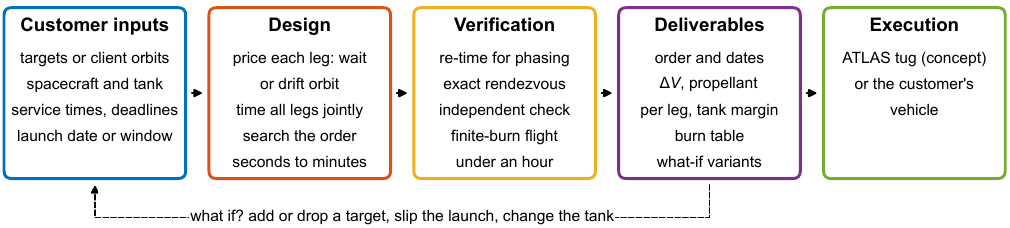}
\caption{Space Postman from the customer's side. Run times are those of
the companion paper~\cite{iskender_iac_c1} for a ten-target tour: design
on one core, verification on six. Deliveries into an orbit need no
along-track phase and are quoted at design fidelity; rendezvous tours are
verified before they are quoted.}
\label{fig:workflow}
\end{figure*}

Figure~\ref{fig:workflow} shows the service from the customer's side. The
customer supplies the targets (catalogue objects, client spacecraft or
client orbits), the spacecraft (dry mass, specific impulse $I_{sp}$,
tank), the service time at each stop and the deadlines or launch window.
The service returns the visiting order, the arrival and departure dates,
the $\Delta V$ and propellant of every leg, the margin left in the tank,
and a burn table with epochs; each what-if change returns a new plan.
For a ten-target tour the design
takes seconds to minutes on one processor
core~\cite{iskender_iac_c1}, and the whole tug study of
Sec.~\ref{sec:tug}, with three independent searches and every baseline,
ran in about ten minutes. Verification of a ten-target rendezvous tour at
the fidelity of the best published reference, including a finite-burn
flight, takes under an hour on a six-core workstation~\cite{iskender_iac_c1}.

\subsection{How the engine works}

The engine follows the multi-fidelity pattern of the best published
multi-target solutions~\cite{b_jpl}. A closed-form screening
model prices each leg in two ways: waiting until the planes come
together, or parking on a drift orbit shaped in altitude, eccentricity
and inclination; it keeps the cheaper. Because a complete tour is priced
in milliseconds, all leg times of a tour can be chosen together, and the
visiting order is searched on top~\cite{iskender_iac_c1,iskender_journal2026}.
For targets that must be met, not just orbits to be reached, the plan is
then re-timed for orbital phasing, solved as an exact rendezvous under
the full oblateness dynamics, checked by an independent propagator and
flown with finite burns in the open-source Basilisk simulator and again
in NASA's GMAT~\cite{b_basilisk,b_gmat,iskender_iac_c1}. Below, results of the
screening model are said to be at \emph{design fidelity}, and results of
the later steps \emph{verified}. The screening model ranks
legs reliably, but its plan costs can move by several percent either way
once the phase is priced; the service therefore quotes verified numbers
for rendezvous tours and treats screening differences of a few percent as
ties~\cite{iskender_iac_c1}. The engine does not choose which targets to
visit when not all fit, offers no optimality certificate, unlike exact
methods for this problem class~\cite{g_leonardi2026,g_bannach2026}, and
models impulsive burns, whereas several peers treat low
thrust~\cite{g_apa2026,g_ciccotti2026}.

\subsection{Business model}

Space Postman is planned as a per-study service in three tiers that
follow the fidelity of Fig.~\ref{fig:workflow}: a quick quote at design
fidelity, a verified plan, and later a flight-ready plan with the
customer's propulsion and navigation models. A seat fee would cover
what-if runs at design fidelity, whose marginal compute cost is small;
the verified tier carries the cost of the high-fidelity runs and of an
analyst's review. Prices are not yet set; Sec.~\ref{sec:value} states the
assumptions used to compare them with building in house.

\section{ATLAS: One Bus, Three Payloads}\label{sec:atlas}

ATLAS is the execution half of the offer: a tug concept, not yet built,
whose payload bay takes one of three cargo classes, with little change
to the bus: de-orbit kits or capture tools for debris removal and
recycling, client spacecraft with release adapters for last-mile
delivery, and propellant with a transfer interface for refuelling. For the
planner the three differ only in what happens at each stop: a kit or a
client leaves the tug, propellant leaves it, or harvested material joins
it~\cite{iskender_iac_c1}. One bus with a cargo-agnostic planner means
one vehicle programme and qualification path instead of one per
business; we see this as the reason to pair the service with a vehicle,
not as a demonstrated cost advantage.

Three operational hooks connect the plan to the vehicle. First, the tug
needs enough autonomy to fly a multi-leg plan without ground
intervention on every leg; the terminal rendezvous and docking at each
target is a separate problem, treated in the first author's earlier
work~\cite{iskender2019dual}. Second,
the propulsion must suit drift transfers: the engine currently plans
impulsive burns, as flown by a storable chemical system, whereas an
electric thruster, possibly including the refuellable thruster that
DEXTER develops~\cite{g_dexter2025}, would need a low-thrust leg
model~\cite{g_apa2026}. Third, the operator needs to replan when a burn is
missed or a target changes. Re-running the design stage from the current
state is fast (Fig.~\ref{fig:workflow}), but in-orbit replanning has not
been demonstrated and is a roadmap item (Sec.~\ref{sec:roadmap}).

\section{Why Operators Can Trust the Plans}\label{sec:trust}

A planning service is only worth buying if its numbers hold when the
mission is flown, so the engine was checked against the best published
solution of a public benchmark. GTOC9 asked for the removal of 123 debris from
Sun-synchronous orbits by a series of missions, one spacecraft per
launch, at the lowest total cost, which grows with the number of
launches and their mass~\cite{b_gtoc9,b_izzo_gtoc9}. The winning solution, by a team
of more than twenty at NASA's Jet Propulsion Laboratory (JPL), used ten
launches, and its mission tables give every leg at full
fidelity~\cite{b_jpl}. Table~\ref{tab:trust} summarises the two-part
check made in the companion paper~\cite{iskender_iac_c1}. First,
calibration: on the winner's published timetable, our solver reproduces
its tenth mission within 0.3\%, and every leg within 7~m/s. Second, the
planner solves the benchmark itself: it takes the same launches, debris
and launch dates as the winner, receives each launch's debris in random
order, and proposes the visiting order, all leg times and every burn.
Verified at full fidelity, its plans need 3.5\% less $\Delta V$ than the
winner's over the ten launches. Eight of the ten missions need less, by
up to 15.6\%; Mission~3, the largest with 21 debris, needs 5.3\% more,
and Mission~9 0.4\% more. On three launches the planner's sequence
happens to coincide with the winner's sequence. On Mission~10 the planner
visits the ten debris in the order 44, 111, 56, 17, 78, 0, 109, 103, 42,
21 in 143.0~days, against the winner's 148.2, with legs of 51 to
299~m/s: the tug rides the node line of each debris while it serves it
and hops where the next line is close. The split of the debris into
launches and the launch dates are a substantial part of the winner's
contribution, so this is a check of the planner, not a ranking; JPL's
post-competition solution, with a partly different split, remains
cheaper~\cite{b_jpl}, as does a ten-launch solution of 698~MEUR at the
maximum launch base cost that team XSCC reported after the competition,
without a validation~\cite{b_xscc}. All ten missions have also been flown
with finite burns, in Basilisk and again in GMAT, a second simulator with
a separate spacecraft model and integrator: they need 0.50\% more $\Delta V$
than with impulsive burns, and the two simulators agree within 12~mm/s on
every mission. The Mission-10 plan needed 1\,416.9~m/s on its first
finite-burn flight, 0.25\% above its impulsive cost. The rule check is
our implementation.

\begin{table}[!htb]
\caption{Credibility check on the benchmark (companion
paper~\cite{iskender_iac_c1}): the published winning
solution~\cite{b_jpl} against our solver on its timetable (calibration)
and against the plans proposed by the Space Postman planner for the same
launches, all at the same fidelity.}
\label{tab:trust}
\centering
\small
\begin{tabular*}{\linewidth}{@{\extracolsep{\fill}}lrrr@{}}
\toprule
\textbf{Check} & \textbf{Winner} & \textbf{This work} & \textbf{Diff.} \\
\midrule
Calibration, Mission 10        & 1\,422.8 & 1\,419.0 & $-0.3\%$ \\
Planner's plan, Mission 10     & 1\,422.8 & 1\,413.4 & $-0.7\%$ \\
Ten missions                   & 25\,136  & 24\,267  & $-3.5\%$ \\
Missions with less $\Delta V$  & --       & 8 of 10  & -- \\
Cost index [MEUR]              & 731.5    & 718.7    & $-1.8\%$ \\
\bottomrule
\end{tabular*}
\par\vspace{3pt}
\begin{minipage}{\linewidth}\footnotesize
$\Delta V$ in m/s: impulsive burns with exact rendezvous under the full
$J_2$ dynamics, independently re-propagated. Calibration: our solver on
the winner's published timetable. Planner: same launches, debris and
launch dates as the winner; the planner proposes every sequence, leg
time and burn. Cost index: the competition's launch-mass cost with a
common 55-MEUR launch base (718.4 against 731.3~MEUR with the official
base costs).
\end{minipage}
\end{table}

\section{Case Study 1: DEXTER, Debris Becomes Propellant}\label{sec:dexter}

\subsection{The case and the question}

DEXTER (Debris Extraction Tools for Extra-terrestrial Recycling) is an
EIC Pathfinder Challenge project under Horizon Europe, from 1~October
2025 to 30~September 2029, with an EU contribution of about 4~MEUR,
coordinated by Cranfield University with Universidad de Alicante, SENER,
Moliri, Magdrive and Vimotek as partners~\cite{g_dexter2025}. It develops robotic tools that harvest
aluminium from derelict hardware and processes that turn the scrap into
propellant for a refuellable thruster, validated in the laboratory and
on the ground; the project itself has no flight phase. Recycling in orbit
has a separate business case~\cite{g_sinharoy2023}, and the
choice of debris by recyclable aluminium content has been studied for
delivery to the Moon~\cite{g_palmer2025}. A mission that applies such
tools in orbit, a notional follow-on to DEXTER, raises the questions a
funder or an operator asks before any hardware is built: which objects,
how many per mission, how long, how much propellant, and how much
harvesting could help. The tour below answers them at design fidelity, and checks the result as an exact rendezvous under the full oblateness dynamics, for a spacecraft that is a stand-in, not a DEXTER design.

\subsection{Why these ten objects}

The candidates are the objects on the published lists of the fifty
statistically most concerning derelicts in
LEO~\cite{g_mcknight2021,g_mcknight2025sdc9}. Rocket bodies dominate these
lists, and the heaviest are the Zenit-2 second stages (SL-16), about nine
tonnes each; eighteen of them share one shell at $71^\circ$ inclination
and 810--860~km mean altitude, a region that ESA's environment report singles
out for its high concentration of collision risk~\cite{g_esa_ser2026}. The shell lies just above the orbits of
Earth-observation satellites developed by ESA, such as Sentinel-5P,
MetOp and FLEX~\cite{iskender_iac_c1}; fragments of a collision there
would spread through that band. For a recycler the cluster has three
further merits: each stage holds an estimated 4.2 to 7.5~tonnes of
aluminium alloy, a bracket between the average aluminium fraction of
upper stages~\cite{g_schulz2026} and the dry mass less the engines; all
eighteen are of one design, so one set of capture and cutting tools fits
every target; and all precess at the same rate.
The last point is also the difficulty: their node offsets are frozen, so
no target ever drifts towards another, and the tug must drift across
them. Of the possible sets of ten stages with consecutive nodes on
1~January~2028, the chosen set has the highest summed collision risk per
unit of estimated $\Delta V$; it spans $135^\circ$ and includes the stages
ranked first on the 2021 and 2025 lists~\cite{iskender_iac_c1}; comparable
sets in the other rocket-body clusters would cost two to three times as
much $\Delta V$.

\subsection{The plan and what it changes}

\begin{figure*}[t]
\centering
\includegraphics[width=\textwidth]{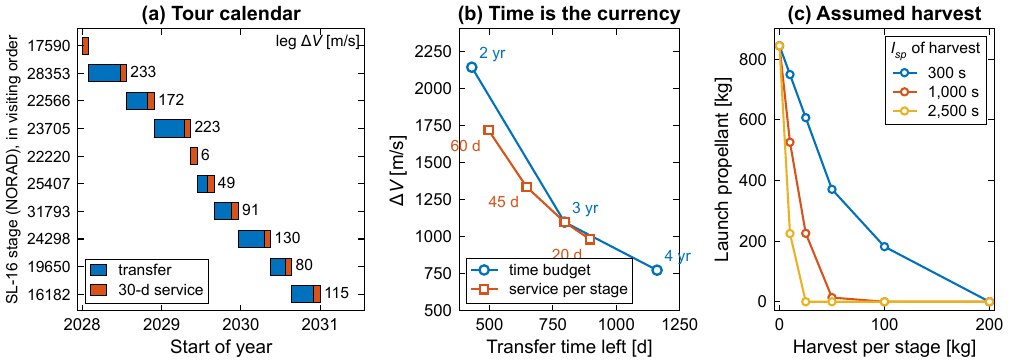}
\caption{DEXTER case study, a notional follow-on tour of ten SL-16
stages (design fidelity, stand-in spacecraft). (a)~Calendar of the
planned tour from 1~January~2028: transfers (blue) and 30-day services
(orange), with the $\Delta V$ of each leg. (b)~$\Delta V$ of the tour
against the transfer time left after servicing, when the time budget
(2, 3, 4 years) or the service time per stage (20--60 days) changes.
(c)~Propellant loaded at launch if the tug turns part of each stage's
aluminium into usable propellant and burns it first, for three assumed
specific impulses; the yields and impulses are parametric assumptions,
not DEXTER performance figures.}
\label{fig:dexter}
\end{figure*}

The tug arrives at the first stage on 1~January~2028, stays 30~days at
each stage for inspection, capture and harvesting (an assumption: there
is no flight precedent for harvesting), and must finish within three
years. The stand-in spacecraft is that of the benchmark in
Sec.~\ref{sec:trust}: 2\,000~kg dry, a 5\,000-kg tank, $I_{sp}=340$~s,
and a 30-kg kit left at each stage~\cite{iskender_iac_c1}.
Figure~\ref{fig:dexter}a shows the resulting calendar. The tour visits
all ten stages in 1\,096~days for 1\,098~m/s and 844~kg of propellant,
a sixth of the stand-in's tank. Every leg but one parks on a drift orbit
50 to 200~km below the shell, which precesses faster and lets the tug's
node fall back across the cluster; the remaining leg is a short hop
between two stages whose nodes have drifted together by then. The walk direction matters: walking the
same ten stages the other way costs 8\% more.

What the plan changes, set against schedules an operator might draw by
hand, is the business content. Waiting alone cannot close a single gap,
because the stages share one rate: without drift orbits the tour would
need about 16~km/s. Splitting the time budget evenly over the legs costs
1\,398~m/s, 27\% more $\Delta V$ and about 290~kg more propellant than
choosing the leg times together. Giving the first legs all the time they want and
the rest what is left overruns the tank.

Above all, time is the currency of this mission
(Fig.~\ref{fig:dexter}b). A two-year budget raises the cost to
2\,142~m/s and 1\,935~kg of propellant; a four-year budget lowers it to
774~m/s and 565~kg. Each further 15~days of service per stage costs 240
to 380~m/s, because it comes out of the drift time; the two sensitivities
fall on nearly one curve against the transfer time left. A funder can
thus trade mission duration against propellant with numbers rather than
rules of thumb, and a harvesting team can see what a faster service is
worth. The floor of the drift orbits matters little: lowering it to
400~km or raising it to 600~km changes the cost by about one percent.

\subsection{Harvested propellant, as a parametric variant}

If the tug turns part of each stage's aluminium into usable propellant
during its service and burns it first, the propellant loaded at launch
shrinks quickly (Fig.~\ref{fig:dexter}c). With 10~kg per stage at
$I_{sp}=1\,000$~s it falls from 844 to 526~kg; with 50~kg per stage at
the same impulse, to 14~kg; at a chemical-like 300~s, 50~kg per stage
still removes more than half. A yield of 50~kg per stage is about one
percent of a stage's estimated aluminium, but these yields are
assumptions, not DEXTER or Magdrive performance figures: the conversion yield, the propellant's specific
impulse and the thruster are what the project is developing. They show
that the tour's logistics leave room for a modest harvest to matter, and
they tell the harvesting team which yields would change the mission.

Two caveats bound this story. The 30-kg kit of the stand-in is not a
removal package for a nine-tonne stage; a real removal solution, not the
tour, would dominate the launch mass. Solved as exact impulsive rendezvous under the full oblateness dynamics, with the tools checked in Sec.~\ref{sec:trust}, the tour costs 1\,118.9~m/s, 1.9\% above its design estimate, with all nine rendezvous closed to within millimetres. Drag on the drift orbits and finite-burn flights are still to come, as is a low-thrust leg model if the tour itself flies
on an electric thruster.

\section{Case Study 2: A Multi-Client Tug}\label{sec:tug}

\subsection{The manifest}

A rideshare launch injects a tug into a 525-km Sun-synchronous orbit
(SSO) with its ascending node at 10:30 local time. The tug carries eight
client spacecraft of 25 to 150~kg, 605~kg in total, and releases each
into its target SSO at 450 to 650~km, with a local time of the ascending node
(LTAN) between 09:30 and 11:30. The manifest is representative, not a real
customer list. The tug is a small storable-propellant carrier (500~kg
dry, a 600-kg tank, $I_{sp}=300$~s); it spends a day at each client,
flies transfers of at most 90~days and must finish within a
year~\cite{iskender_iac_c1}. Drift orbits stay between 400 and 1\,200~km,
above the altitudes where drag would decay them over months. Delivery
into an orbit needs the plane, altitude and eccentricity but no
along-track phase, so the design fidelity is the relevant one here.

\subsection{Why waiting cannot work in SSO}

Every client plane is Sun-synchronous, so every plane precesses at the
Sun's rate and the offsets between clients never change. Waiting closes
nothing: a tug that only waits, or that changes plane at once, needs more
than 6~km/s, almost nine times the propellant its tank holds. Only a
drift orbit, a different altitude or inclination for a while, moves the
tug's plane relative to the clients.

\subsection{What the operator reads}

\begin{figure}[!htb]
\centering
\includegraphics[width=\columnwidth]{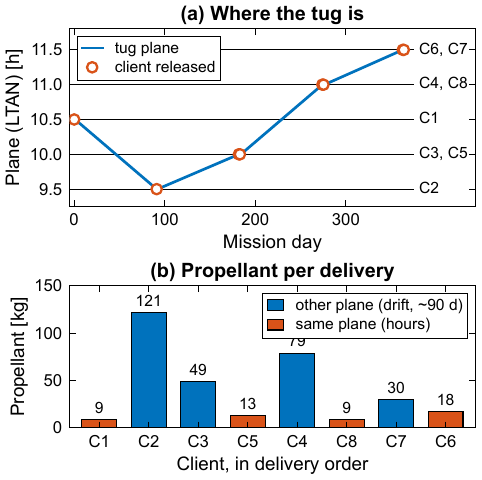}
\caption{Multi-client tug (representative manifest). (a)~Plane of the tug
as the LTAN of its node against mission day; thin black lines are the client
planes and circles the releases. A sloped segment is a drift orbit.
(b)~Propellant spent on the leg that delivers each client, in delivery
order: 327~kg of a 600-kg tank in total.}
\label{fig:tug}
\end{figure}

The plan delivers all eight clients, in five Sun-synchronous planes, in
365~days for 1\,008~m/s and 327~kg of propellant, 55\% of the tank, with
a launch mass of 1\,432~kg. Figure~\ref{fig:tug}a shows the route: the
tug first drifts high (about 720~km) to fall back one hour of LTAN in
90~days, then walks forward through the other planes on lower drift
orbits (438 to 486~km), releasing the clients that share a plane
one after the other within a day. Three independent searches return the
same order. Four levers matter commercially.

\emph{The delivery order.} Delivering the clients in the order in which
the manifest happens to list them needs 2\,385~m/s and 938~kg, more than
the tank holds. A natural hand-made order, sorted by LTAN and then by
altitude, fits, but needs 403~kg, 23\% more propellant than the searched
order. Here the order, not the timing, carries the value: choosing all
leg times together adds almost nothing on this manifest, because the
planes precess together and a delay on one leg does not change the
next~\cite{iskender_iac_c1}.

\emph{Propellant per delivery.} Figure~\ref{fig:tug}b charges each leg's
propellant to the client it delivers. A client in a plane the tug already
occupies costs 9 to 18~kg, only the altitude change; a client half an
hour to an hour of LTAN away costs 30 to 121~kg, and more when the tug is
still heavy early in the mission. This is the basis of a price list by plane offset and by
position in the manifest, rather than by mass alone.

\emph{Clients per tank.} The plan leaves 273~kg of the tank unused. That
margin can be sold as contingency or spent on further clients; whether a
ninth or tenth client fits, and where, is one what-if run.

\emph{Time to orbit.} The propellant-optimal plan releases the last
clients almost a year after launch. A client that needs its orbit sooner
would pay for it in propellant; a delivery deadline per client, which
would let the service price such a request, is one of the planner
extensions of Sec.~\ref{sec:roadmap}.

\subsection{Variant: deploying constellation planes}

The same physics serves a constellation operator. Satellites released
together into a 350-km parking orbit at $53^\circ$ precess about
0.485~deg/day faster than they would in a 550-km shell, so a batch that
waits 41~days before it is raised arrives in a plane $20^\circ$ away from
the first, and a third batch that waits twice as long in a plane
$40^\circ$ away. Each raise costs about
112~m/s, against about 2.1~km/s for turning the plane of the shell by
$20^\circ$ directly. This worked example uses secular $J_2$ drift,
circular orbits and an instantaneous raise. Deployment by differential
precession is established
practice~\cite{g_mcgrath2020,g_dipasquale2022}, and the operational
record matches the example:
Starlink batches have been held at 350~km for 35 and 70~days to reach
planes $20^\circ$ and $40^\circ$ apart~\cite{g_mcdowell2020}. How long
each batch should wait, with several planes, a tank and a deadline for
the whole constellation, is the kind of question Space Postman is built
to answer; we have not yet run the planner on it.

\section{Value Analysis}\label{sec:value}

\subsection{Against realistic baselines}

Table~\ref{tab:value} compares each plan with schedules an operator could
draw without an optimiser: an even split of the time budget, waiting for
the planes instead of using drift orbits, the other natural walk
direction, the manifest order and a hand-sorted order. We deliberately
leave out two other baselines of the companion paper, flying every leg at
once and choosing each leg's time greedily in turn, because no careful
operator would fly them; they illustrate the physics, not the value of a
planner. Against an even split of the time budget or a hand-sorted
order, the plan saves a fifth to a quarter of the propellant, and in
three of the six comparisons it is the difference between a mission that
fits in its tank and one that does not.

\begin{table}[!htb]
\caption{The plans of the two case studies against schedules an
operator could draw by hand (design fidelity, same constraints).}
\label{tab:value}
\centering
\small
\setlength{\tabcolsep}{3pt}
\begin{tabular*}{\linewidth}{@{\extracolsep{\fill}}lrrr@{}}
\toprule
\textbf{Plan} & \textbf{$\Delta V$} & \textbf{$m_p$} & \textbf{vs.\ plan} \\
 & [m/s] & [kg] & [$m_p$] \\
\midrule
\multicolumn{4}{@{}l}{\emph{DEXTER tour, 5\,000-kg tank}} \\
\textbf{Space Postman}       & \textbf{1\,098} & \textbf{844} & -- \\
Even split of the budget     & 1\,398  & 1\,134 & $+34\%$ \\
Other walk direction         & 1\,181  & 910    & $+8\%$ \\
Waiting only                 & 16\,108 & $>10^5$ & over tank \\
\midrule
\multicolumn{4}{@{}l}{\emph{Multi-client tug, 600-kg tank}} \\
\textbf{Space Postman}       & \textbf{1\,008} & \textbf{327} & -- \\
Hand-sorted order            & 1\,175  & 403   & $+23\%$ \\
Manifest order               & 2\,385  & 938   & over tank \\
Waiting only                 & 6\,169  & 5\,391 & over tank \\
\bottomrule
\end{tabular*}
\par\vspace{3pt}
\begin{minipage}{\linewidth}\footnotesize
$m_p$: propellant loaded at launch. Hand-sorted: by LTAN, then altitude.
The DEXTER alternatives other than the walk direction keep the planned
order; the tug's alternative orders are timed with the same optimiser.
Sources: companion paper~\cite{iskender_iac_c1} and its case data.
\end{minipage}
\end{table}

\subsection{What the saving is worth}

For a tug, propellant saved is launch mass saved. The searched order
launches 76~kg lighter than the hand-sorted one and 611~kg lighter than
the manifest order, which would need a larger tank as well. At the
rideshare price reported for 2026, about 7\,000 US dollars per kilogram
beyond a 50-kg base package~\cite{m_spacex_rideshare}, those masses are
worth roughly half a million and four million dollars of launch. For a
debris tour, the roughly 290~kg saved against an even split of the time
budget are about 6\% of the stand-in's tank; with a realistic removal
package per stage, the tour's share of the launch mass shrinks, but its
feasibility does not stop depending on the plan. The larger effect in both
stories is binary: a plan either fits the tank and the deadline or it
does not, and whether it does changes with the order, the timing and the
time budget. The value of servicing has been framed as the value of
flexibility~\cite{g_saleh2003,g_long2007}; a feasible, priced plan within
minutes for every variant a customer asks about makes it quotable.

\subsection{Build or buy}

Table~\ref{tab:buildbuy} lists the cost of building the capability in
house. Apart from the cited
salary anchor and the free availability of NASA's General Mission
Analysis Tool (GMAT)~\cite{b_gmat}, every figure is an assumption for
illustration, not a survey result. On these assumptions an in-house
capability costs a few hundred thousand euros of payroll and one to two
years before its first calibrated plan; a study from the service would
cost a fraction of that and be available at once. The comparison is not
one-sided: an operator with a steady stream of missions may prefer an in-house team, and a single study can also be bought from a consultancy. The
case for the service rests on what the table does not show: the time to
a first calibrated plan, a verification that makes a quote defensible,
and a what-if loop fast enough to answer a customer in one conversation.

\begin{table}[!htb]
\caption{Building a multi-target planning capability in house, against
buying studies. Illustrative; see the Basis column.}
\label{tab:buildbuy}
\centering
\small
\setlength{\tabcolsep}{4pt}
\begin{tabularx}{\linewidth}{@{}Lrl@{}}
\toprule
\textbf{Item} & \textbf{Value} & \textbf{Basis} \\
\midrule
Senior trajectory engineer   & 150--220 kEUR/yr & assumed \\
Mission-analysis engineer    & 90--130 kEUR/yr  & assumed \\
US median, aerospace eng.    & 134\,960 USD/yr  & cited~\cite{m_bls_aero} \\
Time to calibrated planner   & 12--24 months    & assumed \\
Payroll to first plan        & 240--700 kEUR    & derived \\
Propagation tool (GMAT)      & free             & cited~\cite{b_gmat} \\
One study, target price      & 50--200 kEUR     & assumed \\
\bottomrule
\end{tabularx}
\par\vspace{3pt}
\begin{minipage}{\linewidth}\footnotesize
Engineers: one full-time equivalent each; assumed costs are fully loaded
(salary, employer charges, equipment) for Europe. Derived: both engineers
for 12 to 24 months. US median: base pay of aerospace engineers, May
2025, an anchor only. Commercial mission-analysis suites are licensed at
prices that are not public.
\end{minipage}
\end{table}

\subsection{Start bottom-up: software first, assets later}

The offer is meant to grow in four steps of rising capital and time to
revenue, each paying for the next: software that quotes plans as a
service, today; verified plans for partners' missions, made with free
verification tools, which exist today for the benchmark and the DEXTER
tour; plans flown on partners' vehicles, with no hardware to buy; and,
last, a dedicated vehicle, ATLAS, which needs capital and a demonstration
partner. Software-first space companies show that the first steps can
carry a business. Analytical Graphics, the maker of the STK
mission-analysis software, was acquired for US\$700~million in
2020~\cite{m_ansys8k2020}; Exotrail licensed its cloud mission-design
software to Eutelsat in 2020~\cite{m_exotrail_eutelsat2020}, before it
flew its spacevan transfer vehicle (Table~\ref{tab:market}); and Kayhan Space
reported about 20 customers managing more than 500 satellites on
US\$11~million raised~\cite{m_kayhan2023}.

Space Postman is lean by design. A plan takes minutes on one computer
and a verified plan under an hour on a six-core workstation
(Sec.~\ref{sec:postman}); the verification tools, Basilisk and GMAT, are
free~\cite{b_basilisk,b_gmat}; partners fly the plans, so no vehicle has
to be bought before the first customer; and each study is paid for. With
a median of 22~months between a start-up's last raise and its
shutdown~\cite{m_cbinsights2026}, revenue from software before hardware
is the point of this design: a slow market costs months of salaries, not
a vehicle.

\subsection{Risks}

Six risks qualify the business case. (i)~\emph{Market}: in LEO,
multi-target servicing, removal and recycling tours are still concepts
or demonstrations (Table~\ref{tab:market}); demand for tour planning
grows only as those tours are funded. (ii)~\emph{Model}: the
engine plans impulsive burns under the Earth's oblateness alone; drag
during months-long drifts, low-thrust legs and the choice of which
targets to visit are not yet in it. (iii)~\emph{Evidence}: the tug story is at design fidelity with a
representative manifest; the DEXTER tour is verified as impulsive
rendezvous but not yet flown with finite burns, and only the benchmark
plans have been flown with finite burns. (iv)~\emph{Competition}: the
methods are published, including recent drift-orbit servicing
planners~\cite{g_apa2026,g_ciccotti2026,g_leonardi2026}, and
general-purpose mission-analysis suites could add tour planning; the
defensible part is the calibrated, verified workflow and the library of
case studies, not the algorithms. (v)~\emph{Customers}: early customers
are likely to be publicly funded projects, whose timelines are long. (vi)~\emph{Safety}: a derelict Zenit-2 stage may
still carry about 900~kg of oxidiser and 450~kg of
kerosene~\cite{g_odqn28}, so harvesting would first need passivation,
which lengthens the service at each stage and, through the time budget,
raises the propellant of the whole tour.

\section{Roadmap}\label{sec:roadmap}

\paragraph{DEXTER, 2025--2029.} The project runs until September
2029~\cite{g_dexter2025}. Its laboratory results, the harvesting yield,
the conversion efficiency and the thruster's specific impulse and thrust,
are the parameters that Fig.~\ref{fig:dexter}c leaves open; as they
arrive, the tour of Sec.~\ref{sec:dexter} can be re-planned with measured
values, verified at full fidelity and, if the thruster is electric, given
low-thrust legs. The tour is offered as an input to the
mission-concept discussion around DEXTER; it is not a project deliverable.

\paragraph{Product and technology readiness.} The planning engine has
been checked against a published benchmark in simulation and applied to
two further cases, one of them, the DEXTER tour, verified as impulsive
rendezvous. The next step, the aim of a
proposed 24-month development plan, is a prototype that takes
partner-supplied target lists and propulsion data and is validated in
simulated but relevant orbital conditions, which corresponds to TRL~4.
The service tiers follow the evidence: quotes at design fidelity are
possible today, verified plans exist for the benchmark, flown with finite
burns in two simulators, and for DEXTER as impulsive rendezvous, and
flight-ready plans need the customer's propulsion and navigation models.
The model extensions on the roadmap are drag
during long drifts, low-thrust legs, the choice of targets when not all
fit, the split of a whole catalogue into launches and their dates,
delivery deadlines per client, fleets of tugs, and in-orbit replanning with
closed-loop execution under navigation errors and internal disturbances
such as propellant sloshing~\cite{iskender2026slosh}. ATLAS itself is a
concept; flying it requires an in-orbit demonstration partner.

\section{Conclusions}\label{sec:concl}

Debris removal, servicing and refuelling, last-mile delivery, recycling
and constellation deployment are different businesses with one planning
problem underneath: one spacecraft, many targets whose planes drift
apart, one tank and one deadline. Solving it well takes expertise that
most young operators do not have in house, and doing without it is
expensive. Space Postman offers the planning as a service; ATLAS offers
one tug for several cargo classes. Because most start-ups fail when the
money runs out, the offer starts with software that earns from the first
study and adds hardware only when partners and revenue allow.

For a notional follow-on to DEXTER, a tour of ten SL-16 stages above
several ESA-developed Earth-observation satellites closes in three years
for about 1.1~km/s with a stand-in spacecraft, where waiting alone cannot
close a single gap and an even split of the time budget costs a third
more propellant; the time budget is the currency of such a mission. For a
tug with eight clients from one rideshare, the delivery order decides
whether the mission fits the tank, and the plan prices each delivery by
the plane it must reach. The engine reproduces the published winning
solution of a public debris-removal benchmark within a fraction of a
percent at that solution's fidelity and, given the same launches, debris
and launch dates, proposes plans that need 3.5\% less $\Delta V$ and
that fly with finite burns, in two independent simulators, for half a
percent more.

The tug case is at design fidelity and the DEXTER tour verified only as impulsive rendezvous; the model ignores drag, and in-orbit replanning is not yet demonstrated. Read at
the fidelity at which it was obtained, the evidence supports the claim of
this paper: a calibrated planner, paired with a payload-agnostic tug,
lowers the barrier that keeps new operators out of orbital logistics.

\section*{Acknowledgements}
The authors thank Final Proximity Space Systems for its support, and ESA
for organising the benchmark competition of Sec.~\ref{sec:trust} and
making its data public. The orbital elements of the DEXTER candidates
are public elements from CelesTrak.
The relationships of Final Proximity Space Systems and Magdrive to this
work are disclosed in Sec.~\ref{sec:market}.

\bibliographystyle{unsrt}
\bibliography{references}

\end{document}